\documentclass[twocolumn]{aastex631}

\usepackage{hyperref}
\usepackage{float}
\usepackage{threeparttable}
\usepackage{amsmath}

\usepackage{amsmath}
\usepackage{amssymb}
\usepackage{xspace}
\usepackage{xifthen}
\usepackage{eso-pic}

\newcommand{\Gaia}{{\it Gaia}\xspace}

\newcommand{\secref}[1]{Section~\ref{sec:#1}}

\newcommand{\tabref}[1]{Table~\ref{tab:#1}}
\newcommand{\figref}[1]{Figure~\ref{fig:#1}}
\newcommand{\eqnref}[1]{Equation~\eqref{eqn:#1}}

\mathchardef\mhyphen="2D
\newcommand{\roughly}{\ensuremath{ {\sim}\,} }
\newcommand{\code}[1]{\texttt{#1}\xspace}
\newcommand{\Sin}{\mathrm{sin}\xspace}
\newcommand{\Cos}{\mathrm{cos}\xspace}

\newcommand{\unit}[1]{\ensuremath{\mathrm{\,#1}}\xspace}
\newcommand{\uni}[1]{\ensuremath{\mathrm{#1}}\xspace}

\newcommand{\masy}{\unit{mas~yr^{-1}}}
\newcommand{\mas}{\unit{mas}}
\newcommand{\masmag}{\unit{mas~mag^{-1}}}
\newcommand{\asec}{\uni{\arcsec}}

\newcommand{\magn}{\unit{mag}}

\newcommand{\bandvar}[2][]{%
  \ifthenelse{\isempty{#1}}{\var{#2}}{\var{#2\_#1}}%
}

\providecommand{\ZD}{\ensuremath{\mathrm{ZD}}}
\providecommand{\mra}{\ensuremath{\alpha}}
\providecommand{\mdec}{\ensuremath{\delta}}

\newcommand{\var}[1]{\ensuremath{\texttt{\MakeUppercase{#1}}}\xspace}

\newcommand{\SExtractor}{\code{SExtractor}}
\newcommand{\sextractor}{\SExtractor}
\newcommand{\PSFEx}{\code{PSFEx}}
\newcommand{\scamp}{\code{SCAMP}}

\newcommand{\PMO}{\affiliation{Purple Mountain Observatory, Chinese Academy of Sciences, Nanjing 210023, China}}
\newcommand{\USTC}{\affiliation{School of Astronomy and Space Sciences, University of Science and Technology of China, Hefei 230026, China}}
\newcommand{\TDLI}{\affiliation{Tsung-Dao Lee Institute and State Key Laboratory of Dark Matter Physics, Shanghai Jiao Tong University, Shanghai, 201210, China;}}
\newcommand{\DAUSTC}{\affiliation{Department of Astronomy, University of Science and Technology of China, Hefei 230026, China}}
\newcommand{\IDSS}{\affiliation{Institute of Deep Space Sciences, Deep Space Exploration Laboratory, Hefei, 230026, China}}

\newcommand{\LabPDE}{\affiliation{State Key Laboratory of Particle Detection and Electronics, University of Science and Technology of China, Hefei 230026, China}}

\newcommand{\HIT}{\affiliation{School of Aerospace Information, Hefei Institute of Technology, Zhizhong Road, Hefei, Anhui 238706, China}}

\shorttitle{Astrometry Calibration with Zernike Polynomials for wide-field images}
\shortauthors{Yang et al.}

\begin{document}

\title{WFST Astrometric Calibration---I. Modeling Global Geometric Distortion with Zernike Polynomials}


\author{Chao Yang}
\PMO
\USTC

\author{Min Fang}
\PMO
\USTC

\author{Xian~Zhong~Zheng}
\TDLI
\PMO                                 

\author{Guoliang Li}
\PMO
\USTC

\author{Binyang Liu}
\PMO

\author{Zheng Lou}
\PMO

\author{Zhen Wan}
\DAUSTC
\USTC

\author{Miaomiao Zhang}
\PMO

\author{Tian-Rui~Sun}
\PMO

\author{Lulu Fan}
\DAUSTC
\USTC
\IDSS

\author{Xiaoling Zhang}
\PMO

\author{Xu Kong}
\DAUSTC
\USTC
\IDSS

\author{Yongquan Xue}
\DAUSTC
\USTC

\author{Wen Zhao}
\DAUSTC
\USTC

\author{Bin Li}
\PMO

\author{Wentao Luo}
\HIT

\author{Feng Li}
\LabPDE

\author{Wei Liu}
\PMO

\author{Jian Wang}
\LabPDE
\IDSS

\author{Hongfei Zhang}
\LabPDE

\author{Hao Liu}
\LabPDE

\author{Qinfeng Zhu}
\DAUSTC
\USTC
\IDSS

\author{Hairen Wang}
\PMO

\author{Dazhi Yao}
\PMO


\correspondingauthor{Xian~Zhong~Zheng}
\email{xzzheng@sjtu.edu.cn}

\begin{abstract}

Accurate modeling of geometric distortion is essential for precise astrometric calibration in wide-field imaging surveys. We present a self-calibration method based on Zernike polynomials, applied to imaging data from the Wide Field Survey Telescope (WFST). Our approach constructs a global geometric distortion (GD) model from the position offsets of stars in the WFST $r$-band relative to \Gaia DR3, achieving a median systematic uncertainty of below 10\mas for individual exposures. The correspondence between Zernike polynomials and optical aberrations reveals that the global GD of WFST is dominated by coma, inherent to the optical design, while rapid variations are likely attributed to the atmospheric dispersion corrector. Applying this method to 82 exposures from a single night (20250218), we find that the relative positions of the WFST CCDs remain stable, with standard deviations of less than 0.1\,pixel in translation and  $1\farcs8$ in rotation. The corrected WFST astrometric system is thereby tied to the \Gaia DR3 coordinate frame, with further refinements to be presented in future work.

\end{abstract}

\keywords{Astrometry -- Astronomy image processing}

\section{Introduction} 
\label{sec:intro}

With the advent of large-scale optical sky surveys, such as the Sloan Digital Sky Survey (SDSS; \citealt{York2000}), the superior accuracy and stability of Charge-Coupled Device (CCD) cameras over photographic plates have enabled increasingly precise and rapid measurements of astronomical sources over wide areas of the sky. At the same time, modern telescopes equipped with wide-field cameras have further enhanced the quality and efficiency of observations. Of them, the Canada-France-Hawaii Telescope (CFHT), a 3.6\,m telescope equipped with MegaCam\citep{Boulade2003}, offers a field of view (FoV) of $\sim 1\unit{deg^2}$. The Dark Energy Camera (DECam; \citealt{Flaugher2015}), mounted on the Blanco 4\,m telescope, provides a wider FoV of $\sim 3\unit{deg^2}$ and and has supported deep and wide-area optical surveys, most notably the Dark Energy Survey (DES; \citealt{des2005}). The Hyper Suprime-Cam (HSC; \citealt{Aihara2018}) on the 8.2\,m Subaru Telescope further extends this capability by combining a FoV of $\sim 1.8\unit{deg^2}$ with excellent image quality, enabling deep, high-resolution wide-field imaging. The Zwicky Transient Facility (ZTF) Observing System \citep{Dekany2020}, which installed on the 48-inch Samuel Oschin Telescope \citep{Bellm2019} at the Palomar Observatory, provides a $47\unit{deg^2}$ FoV, enabling high-cadence monitoring of time-domain and transient phenomena within only two nights. And the Rubin Observatory Legacy Survey of Space and Time (LSST), with a $9.6\unit{deg^2}$ FoV \citep{Ivezi2019,rubin2025}, will survey $\sim 20,000\unit{deg^2}$ of the southern sky to a depth of $r \approx 27.5$\magn. Nevertheless, observations with ground-based telescopes are still influenced by atmospheric effects \citep{Fortino2021, White2022}, making further improvements in astrometric and photometric calibration an ongoing challenge. 

\Gaia DR3 \citep{Lindegren2021} provides accurate astrometry for 585 million sources with five-parameter solutions, with median positional uncertainties ranging from $\sim0.012$\mas ($G=13$\magn) to $\sim0.374$\mas ($G=20$\magn). This unprecedented precision offers a reliable reference frame that significantly reduces uncertainties in astrometric calibration. This enables accurate self-calibration even at the single-CCD level, achieving tens of milliarcsecond accuracy \citep{Wan2019}. However, single-CCD calibration faces key limitations. First, the reduced number of reference stars can lead to insufficient sampling for some corrections (e.g., differential chromatic refraction (DCR) and lateral color corrections), thereby introducing additional errors. Second, geometric distortion (GD), arising from the telescope's optical design, affects the entire focal plane and leads to systematic position offsets in measured source positions. Multi-CCD calibration can model the global GD \citep{Anderson2003,Bernstein2017,Magnier2020,Zheng2022} and also provides a quantitative diagnostic of the telescope's optical design and operational state.

Astrometric calibration is an indispensable part of image processing. Although \Gaia DR3 provides an unprecedentedly precise reference catalog, its depth ($G \approx 20$\magn) is often insufficient for calibrating the observations taken with telescopes of a small FoV (e.g., the Hubble Space Telescope and the James Webb Space Telescope), in which the number of unsaturated stars can be limited \citep{Whitmore2016, Libralato2024}. In addition, since stellar positions from \Gaia DR3 are given at epoch J2016, propagating them to the epoch of the observations introduces additional errors that increase with epoch difference. Over long intervals, the uncertainties from the proper motions of sources can become the dominant contribution to the final positional uncertainties of the reference sources. Though \Gaia's proper motion precision is also unparalleled (median uncertainties of $\sim0.016\masy$ at $G=13$\magn and $\sim0.487\masy$ at $G=20$\magn; \citealt{Lindegren2021}), high-accuracy astrometric calibration ($\lesssim10\mas$) still suffers from the reduced number of reference stars, particularly in crowded fields, where the uncertainties of proper motions are usually larger. These caution the increasing uncertainties in using fainter stars for astrometric calibration.

High-accuracy astrometric calibration for ground-based telescopes has been actively advanced in recent years. \cite{Lubow2021} and \cite{White2022} corrected the PS1 DR2 \citep{Magnier2020} for GD and DCR, achieving a median uncertainty of $\sim5$\mas at $i = 17$\magn. Similarly, \cite{Bernstein2017} corrected GD, tree rings \citep{Plazas2014}, DCR and lateral color effects in Dark Energy Survey (DES) images, improving the astrometric accuracy to 3$-$6\mas for individual exposures. Subsequently, \cite{Fortino2021} applied Gaussian process regression to futher mitigate astrometric uncertainties induced by atmospheric turbulence. In this work, we present a global astrometric calibration framework based on the first year of Wide Field Survey Telescop (WFST, \citealt{Wang2023}) data. This framework is designed to serve as a foundation for implementing advanced corrections, including those for DCR, lateral color, Charge Transfer Efficiency (CTE), and Gaussian process interpolation. Furthermore, by quantifying global GD will allow us to assess the astrometric accuracy of WFST data products and to monitor the telescope's optical performance and operational stability. To this end, we construct a global GD model using Zernike polynomials.

The paper is organized as follows: \secref{zernike} introduces the mathematical formulation, properties, and applications of Zernike polynomials; \secref{data} describes the WFST and the datasets; \secref{algorithm} presents the process of constructing the GD model; \secref{result} shows the WFST GD pattern and examines the stability of the GD model between different exposures. We also discuss the astrometric accuracy and relative positions of CCDs in this section. Finally, we conclude in \secref{sum}.

\section{Zernike Polynomials}
\label{sec:zernike}

\subsection{Mathematical Properties}
\label{sec:zproperty}
The zernike polynomials are a set of orthogonal functions defined over the unit disk \citep{Zernike1934, Bhatia1954}. They are named after the optical physicist Frits Zernike, who was awarded the Nobel Prize in 1953. Their mathematical form is given as follows:
\begin{equation}
\footnotesize
     \Phi_j(x,y) = Z_n^m(\rho, \theta) = N_n^m \, R_n^m(\rho) \, \Theta_m(\theta), \\
\end{equation}

\noindent where $\rho$ and $\theta$ denote the polar coordinates on the unit disk, with
\begin{equation}
\footnotesize
\begin{cases}
     \rho = \sqrt{x^2 + y^2},\quad  0 < \rho < 1, \\[4pt]
     \theta = arctan(\frac{y}{x}),\quad  0 < \theta < 2\pi, \\[4pt]
     |Z_n^m(\rho, \theta)| < 1. \\
\end{cases}
\end{equation}

\noindent Here, the normalization coefficient is defined as
\begin{equation}
\footnotesize
     N_n^m = \sqrt{(2-\delta_{0, m})\,(n+1)},
\end{equation}

\noindent the radial coefficient is defined as
\begin{equation}
\footnotesize
\begin{split}
     R_n^{|m|}(\rho) = \sum_{k=0}^{\frac{(n - |m|)}{2}}(-1)^k \, \frac{(n - k)!}{k! \left( \frac{n + |m|}{2} - k \right)! \, \left( \frac{n - |m|}{2} - k \right)!} \, \rho^{n - 2k},
\end{split}
\end{equation}

\noindent and the radian expression is defined as
\begin{equation}
\footnotesize
     \Theta_m = 
     \begin{cases}
          \Cos(m\theta),\quad m \geq 0, \\[4pt]
          \Sin(|m|\theta),\quad m < 0.\\
     \end{cases}
\end{equation}

\noindent The index $j$ represents the Noll index \citep{Noll1976}, which is used to convert the $m$ and $n$ indices in the application into a single index. The transformation between them is defined as follows:
\begin{equation}
\footnotesize
     j = \frac{n \, (n+1)}{2} + |m| + 
     \begin{cases}
          0,\quad m > 0 \land n \equiv {0,1}\,(\mathrm{mod}\,4), \\[4pt]
          0,\quad m < 0 \land n \equiv {2,3}\,(\mathrm{mod}\,4), \\[4pt]
          1,\quad m \geq 0 \land n \equiv {2,3}\,(\mathrm{mod}\,4), \\[4pt]
          1,\quad m \leq 0 \land n \equiv {0,1}\,(\mathrm{mod}\,4). \\
     \end{cases}
\end{equation}

When applied to practical problems such as our astrometric solution, Zernike polynomials exhibit several mathematically distinct features from other polynomials. Foremost among these is orthogonality \citep{Mahajan2007}, with

\begin{equation}
     \label{eqn:ortho}
     \footnotesize
          \int_{0}^{2\pi} \int_{0}^{1} Z_n^m(\rho, \theta) \, Z_{n'}^{m'}(\rho, \theta) \, \rho d\rho d\theta 
          = \frac{\pi}{1 + \delta_{m0}} \cdot \frac{1}{n + 1} \cdot \delta_{nn'} \cdot \delta_{mm'}.
     \end{equation}

\noindent It ensures that polynomials of different orders are independent, allowing them to effectively represent various planar wavefronts or images. Accordingly, it can be expressed as

\begin{equation}
     \label{eqn:wavefront}
     \footnotesize
          W(x, y) =  \sum_{j=1}^{J} a_j \, Z_j(x, y).
      \end{equation}

Furthermore, by definition, Zernike polynomials are applicable only over the unit disk. However, the sources used for our astrometric calibration do not lie within a unit disk due to the CCD layout (\figref{fov}) and the optical design. As a result, constructing the astrometric solution inevitably requires translating, rotating, and resizing the Zernike polynomials. In the following, we illustrate how predefined Zernike polynomials are affected by these operations \citep{Bar2006, Lundstr2007}.

We first consider the case of translation. Let $\Delta x$ and $\Delta y$ represent the displacements along the $x$ and $y$ axes in \eqnref{wavefront} \citep{Wang1980}, respectively. Following \citet{Guirao2001} and \citet{ANSI2017}, expanding $W(x,y)$ in a Taylor series with respect to these displacements gives

\begin{equation}
\label{eqn:shift}
\footnotesize
     \begin{aligned}
          & W(x-\Delta x, y-\Delta y) \\[4pt]
          & \quad =  \sum_{i=0}^{\infty} \, \frac{(-1)^i}{i!} \, (\Delta x \frac{\partial}{\partial x} + \Delta y \frac{\partial}{\partial y})^i \, W(x, y) \\[4pt]
          & \quad = \sum_{n, m}^{} a_n^m \sum_{i=0}^{\infty} \, \frac{(-1)^i}{i!} \, (\Delta x \frac{\partial}{\partial x} + \Delta y \frac{\partial}{\partial y})^i\, Z_n^m(x, y),
     \end{aligned}
\end{equation}

\noindent where

\begin{equation}
\footnotesize
     \begin{aligned}
         & \frac{\partial Z_n^m(x, y)}{\partial x} \\[4pt]
         & \quad = (1+\delta_{m0}) \left[ \sum_{n'=|m|+1}^{n-1}(n'+1) \, {Z_{n'}}^{\frac{m}{|m|}(|m|+1)}\right. \\[4pt]
          & \quad \left. + (1-\delta_{m0}) \, (1-\delta_{m-1})\sum_{n'=|m|-1}^{n-1}(n'+1) \, {Z_{n'}}^{\frac{m}{|m|}(|m|-1)} \right]   \\[4pt]
\end{aligned}
\end{equation}

\noindent and 

\begin{equation}
     \footnotesize
          \begin{aligned}
         & \frac{\partial Z_n^m(x, y)}{\partial y} \\[4pt] 
         & \quad = (1+\delta_{m0})\frac{m}{|m|} \left[ \sum_{n'=|m|+1}^{n-1}(n'+1) \, {Z_{n'}}^{-\frac{m}{|m|}(|m|+1)}\right. \\[4pt]
          & \quad \left. - (1-\delta_{m0}) \, (1-\delta_{m1})\sum_{n'=|m|-1}^{n-1}(n'+1) \, {Z_{n'}}^{-\frac{m}{|m|}(|m|-1)} \right].   \\
     \end{aligned}
\end{equation}

Next, consider applying a rotation of the Cartesian coordinate system by an angle $\alpha$ about the origin. Incorporating this rotation into \eqnref{shift} leads to the updated Zernike polynomial coefficients, denoted as $b_{n}^{m}$, given in \citet{Niu2022} as

\begin{equation}
\label{eqn:rotating}
\footnotesize
     b_{n}^{m} = -a_{n}^{-m}\Sin(m\alpha)+a_n^{m}\Cos(m\alpha).
\end{equation}

In the case of a resizing transformation, let $W_1$ and $W_2$ represent the Zernike polynomials before and after resizing, respectively. Denote their corresponding maximum radii as $R_1$ and $R_2$ ($R_2 \leq R_1$), and define the radius ratio as $\epsilon$. Under these definitions, we obtain
\begin{equation}
\label{eqn:resize}
\footnotesize
     \begin{aligned}
          W_2(R_2 \rho, \theta) & = \sum_{j=0}^{J} b_j Z_j(\rho, \theta), \\[4pt]
          W_2(R_2 \rho, \theta) & = W_1(\epsilon R_1 \rho, \theta) = \sum_{j=0}^{J}a_j Z_j (\epsilon \rho, \theta).\\[4pt]
     \end{aligned}
 \end{equation}

\noindent From \eqnref{resize}, \cite{Dai2008} derived
\begin{equation}
\label{eqn:resizing}
\footnotesize
     \begin{aligned}
          b_j = b_n^m = & \sum_{i=0}^{[(n_{max}-n)/2]} G_n^i (\epsilon) a_{n+2i}^m,
     \end{aligned}
\end{equation}

\noindent where
\begin{equation}
     \footnotesize
          \begin{aligned}
               G_n^i (\epsilon) a_{n+2i}^m & = \sqrt{(n+2i+1)(n+1)} \\[4pt]
               & \times\left[ \sum_{j=0}^{i}\frac{(-1)^{i+j}(n+i+j)!}{j!(n+j+1)!(i-j)!}\epsilon^{2j}\right]\epsilon^n.
     \end{aligned}
\end{equation}

By combining \eqnref{shift}, (\ref{eqn:rotating}), and (\ref{eqn:resizing}), it is clear that applying translation, rotation, and resizing to a set of $n$th-order Zernike polynomials allows the original function to be reconstructed using higher-order polynomials \citep{Guirao2001,Niu2022,Dai2008}. Therefore, when using Zernike polynomials to derive astrometric solution for our images, the inclusion of higher-order terms is necessary. These terms not only capture higher-order optical distortions but also compensate for the loss of accuracy introduced by applying translation, rotation, and resizing to lower-order Zernike polynomials in the GD modeling.

\subsection{Application}
\label{sec:app}
The unique characteristics of Zernike polynomials have enabled their widespread application in various fields (figure~28 of \citealt{Niu2022}), including optical design \citep{Ferreira2015}, diffraction theory, and adaptive optics (AO; \citealt{Schock2003}). Ideally, the wavefront of starlight arrives at a telescope as a plane. However, variations in air temperature, humidity and  density in the atmosphere alter the refractive index, and turbulence within the atmospheric layers induces perturbations, resulting in phase differences across the wavefront \citep{Davies2012}. In an AO system, such aberrations are measured by a wavefront sensor (e.g., a Shack-Hartmann sensor), and Zernike polynomials are used to reconstruct the entire wavefront error map. The deformable mirror then compensates the incoming light based on this map to improve the overall imaging resolution. In fact, Zernike polynomials of different orders correspond well to different types of aberrations \citep{Mouroulis1997}, as shown in \tabref{zernike}, making them particularly suitable for identifying and characterizing optical aberrations.

For our astrometric solution, inspired by the framework used in the AO system, we characterize the distortion at each location on the focal plane by the offset between the measured and true positions of stars. Using offsets from stars distributed across the entire focal plane, we model the distortion with Zernike polynomials to construct a global compensation map. Notably, in telescopes equipped with mosaic CCD cameras, the distortion introduced by the optical design is continuous across the focal plane. As a result, the astrometric solution derived from this approach naturally incorporates the relative positions of the individual CCDs.

\begin{table}
     \setlength{\tabcolsep}{8pt}
     \begin{center}
         \caption{Correspondence between Zernike polynomials and common optical aberration modes}
         \label{tab:zernike}
         \begin{tabular}{llrl}
         \hline
         \hline
            j (Noll) & n & m & Aberration name\\
         \hline
            1 & 0 & 0 & Piston \\
         \hline
            2 & 1 & $+1$ & Vertical Tilt \\
            3 & 1 & $-1$ & Horizontal Tilt \\
         \hline
            4 & 2 & 0 & Defocus \\
            5 & 2 & $-2$ & Oblique Astigmatism \\
            6 & 2 & $+2$ & Vertical Astigmatism \\
          \hline
            7 & 3 & $-1$ & Vertical Coma \\
            8 & 3 & $+1$ & Horizontal Coma \\
            9 & 3 & $-3$ & Horizontal Trefoil \\
            10 & 3 & $+3$ & Oblique Trefoil \\
          \hline
            11& 4 & 0 & Primary Spherical \\
            12& 4 & $+2$ & Vertical Secondary Astigmatism\\
            13& 4 & $-2$ & Oblique Secondary Astigmatism \\
            14& 4 & $+4$ & Vertical Quadrafoil \\
            15& 4 & $-4$ & Oblique Quadrafoil \\
         \hline
         \end{tabular}
     \end{center}
\end{table}

\section{Instrument and Data}
\label{sec:data}

The WFST is an optical telescope with a FoV of 3\,deg diameter \citep{Wang2023}. Its optical system consists of a 2.5\,m primary mirror and a prime-focus assembly, consisting of five corrector lenses, an atmospheric dispersion compensator (ADC), and a science camera equipped with mosaic detectors of nine CCDs (E2V CCD290-99), each with $9216 \times 9232$ pixels and a pixel size of 10\unit{\mu m}.

Benefiting from the wide FoV of WFST and its pixel scale of $0\farcs332$, in the $r$ band, even 30\,s exposure provides a sufficiently large number of stars to determine simultaneously for the global GD model and the relative positions of the CCDs. Therefore, our dataset consists of all useful $r$-band exposures obtained from eight observing nights, as summarized in \tabref{data}. 

\begin{table*}[t]
     \setlength{\tabcolsep}{10pt}
     \begin{center}
     \footnotesize
         \caption{WFST observing sequences}
         \label{tab:data}
         \begin{tabular}{lcllll}
         \hline
          Epoch$^{a}$ & Field & ExpCount & ExpTime (s)& Airmass & PSF FWHM (\asec)\\
         \hline
          20240215 & DHugr\_s27, NEP, COSMOS, N0xxxx, 2016HO3 & 122 & $30 \sim 180$ & $ 1.08 \sim 1.38$ & $1.32 \sim 3.39$\\
          20240327 & DHugr\_s27, NEP, COSMOS, N0xxxx, EP240222a & 121 & $60 \sim 180$ & $ 1.05 \sim 1.97$ & $1.02 \sim 3.35$\\
          20240513 & DHugr\_s27/s15, NEP, N0xxxx, Boötes I/III & 127 & $60 \sim 180$ & $ 1.12 \sim 1.92$ & $1.18 \sim 1.98$\\
          20250129 & DHugr\_a36, NEP, S00105 & 77 & $60 \sim 180$ & $ 1.27 \sim 1.89$ & $0.95 \sim 1.76$\\
          20250206 & DHugr\_a36, WFsci, N00xxx & 86 & $30 \sim 60$ & $ 1.01 \sim 1.96$ & $0.92 \sim 1.78$\\
          20250218 & DHugr\_a36, WFsci, S250206, N00301 & 142 & $30 \sim 75$ & $ 1.01 \sim 2.18$ & $1.08 \sim 2.22$\\  
          20250219 & DHugr\_a36/s36, WFref, WFsci & 159 & $30 \sim 90$ & $ 1.07 \sim 1.97$ & $0.98 \sim 2.09$\\
          20250318 & DHugr\_s27, WFsci, WFref, NEP, N0xxxx & 111 & $30 \sim 90$ & $ 1.05 \sim 2.20$ & $0.75 \sim 1.84$\\
         \hline
         \end{tabular}
     \end{center}
     \begin{flushleft}
          \footnotesize \textbf{Note.} $^{a}$ The date at the start of the night in CST.
     \end{flushleft}
\end{table*}

\section{Methodology}
\label{sec:algorithm}

\subsection{Source Centroid Determination}
\label{sec:centroid}
Following the general framework outlined in \secref{app}, accurately determining the position $(X,Y)$ of one source on CCD is a fundamental step in our processes. Using observational data, \cite{Li2009} compared three commonly used centroiding algorithms--Modified Moment, Two-dimensional Gaussian Fit, and Median. And they found that Two-dimensional Gaussian Fit provides higher centroiding precision. However, When stellar images are undersampled (i.e., FWHM $<2$\,pixels) or when faint stars exhibit inherently low pixel-level signal-to-noise ratios, the effective point spread function (ePSF) fitting method provides comparable centroiding precision to Two-dimensional Gaussian Fit \citep{Anderson2000}, while significantly reducing the adverse effects of these limitations \citep{Bellini2010, Lin2021}.

After applying bias and flat calibrations for each exposure, we use \sextractor and \PSFEx to model the PSF and obtain PSF-based positions for all detected sources. In the reference catalog, the renormalised unit weight error (\code{ruwe}) provided by \Gaia DR3 is expected to be close to 1.0 for sources that are well-described by a single-star model, so non-stellar sources are excluded using the criterion $\code{ruwe} > 1.4$ to avoid systematic centroid offsets between extended sources observed from space and the ground telescopes. Since stellar positions from \Gaia DR3 are given in the epoch of J2016 and our observations were obtained more than nine years later. We update the positions of references stars to the epoch when WFST observations were taken, using \Gaia proer motions. To pursue an astrometric accuracy to a level of a few milliarcseconds, we further remove reference stars with large uncertainties in proper motion with the critera of $\code{pmra\_err} > 0.2\masy$ or $\code{pmdec\_err} > 0.2\masy$, as well as sources with measured position uncertainties exceeding $5\mas$ as estimated by \sextractor.

\subsection{Star Matching}
\label{sec:match}

An initial fourth-order TAN-PV world coordinate system (WCS) is constructed using \scamp to transform $(X, Y)$ coordinates into the celestial coordinates  $(\mra, \mdec)$ for all detected sources. A preliminary photometric calibration at the CCD level is then carried out with the WFST synthetic photometric catalog, which is generated from \Gaia BP/RP (XP) spectra \citep{De2023,Montegriffo2023}. And saturated sources, identified as those with measured magnitudes significantly fainter than the reference values at the bright end, are excluded from this process. Then, initial source matching is performed by associating each detected source with the nearest reference star within a tolerance of 10\asec. 

For ground-based observations, atmospheric effects can introduce biases that hinder accurate source matching. First, seeing-limited resolution causes blended stellar profiles on the CCD, biasing centroid measurements. Second, atmospheric extinction reduces detected stellar fluxes and varies with time, complicating the use of brightness as a matching constraint. Therefore, for non-variable stars, a per-exposure transformation between \Gaia $G$ band and WFST $r$ band is required to achieve robust photometric consistency and secure star matching.

We correct for photometric system differences and atmospheric effects by fitting the following relation to the matched stars: 

\begin{equation}
\label{eqn:band}
\footnotesize
     r = a_0 + a_1 \times G + a_2 \times (BP-RP) + a_3 \times (BP-RP)^2. 
\end{equation}

\noindent Then a sigma-clipping ($3\sigma$) procedure is applied to the residuals between the transformed and measured magnitudes to retain stars that are consistently matched in both position and brightness. Finally, stars with brighter neighbors within 10\asec are removed, yielding a clean catalog of matched stars for constructing the global GD of each exposure.

\begin{figure}
     \center
     \includegraphics[width=\columnwidth]{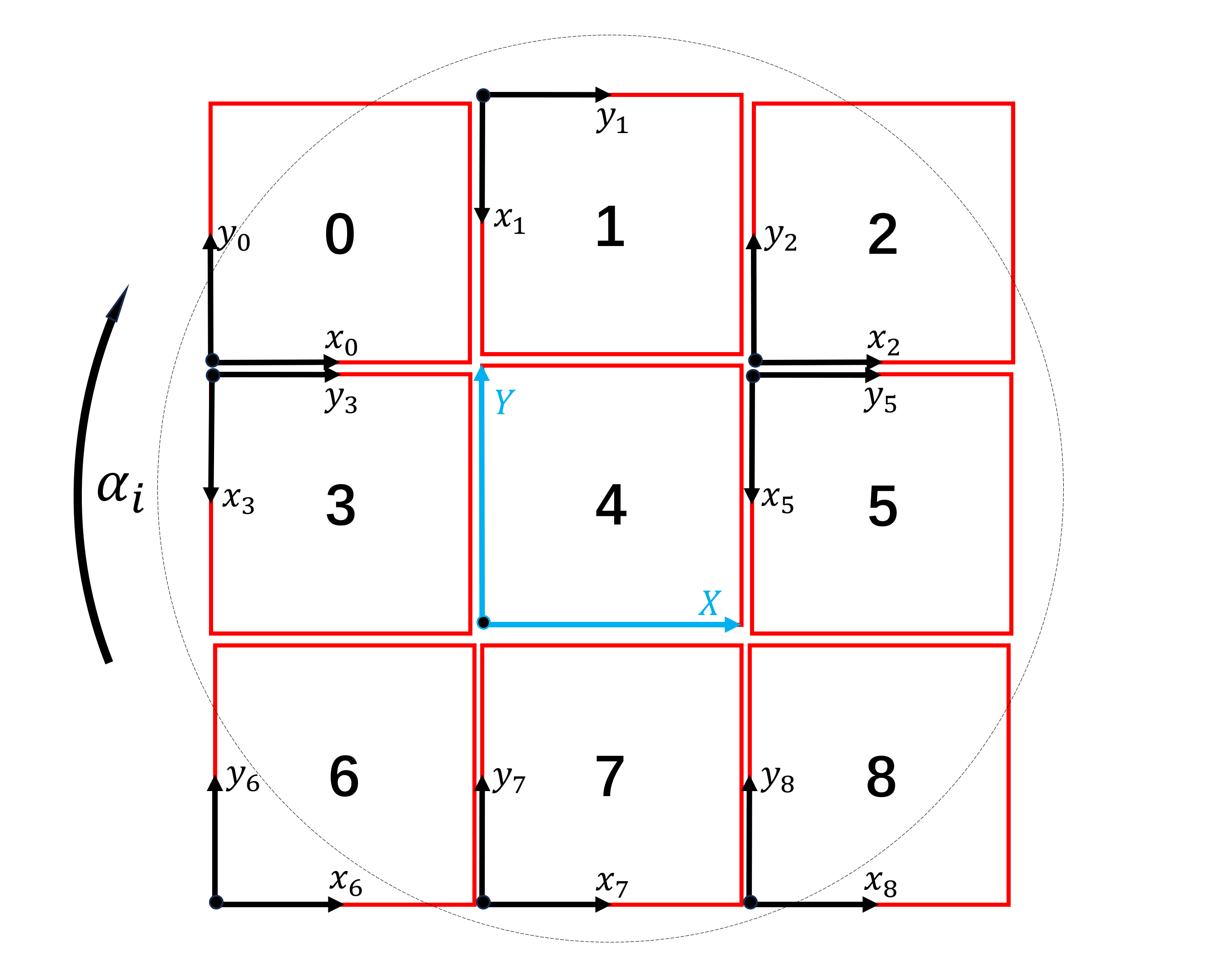}
     \caption{The mosaic CCD layout of WFST. The light gray circle marks the FoV of WFST, while the red rectangles represent individual CCDs. Black coordinate axes denote the local Cartesian system of each CCD, and the blue axes indicate the global Cartesian system of the FoV. The parameter $\alpha_i$ represents the rotation angle of each CCD relative to the central CCD.}
     \label{fig:fov}
\end{figure}

\begin{figure*}[t]
     \center
     \includegraphics[width=\textwidth]{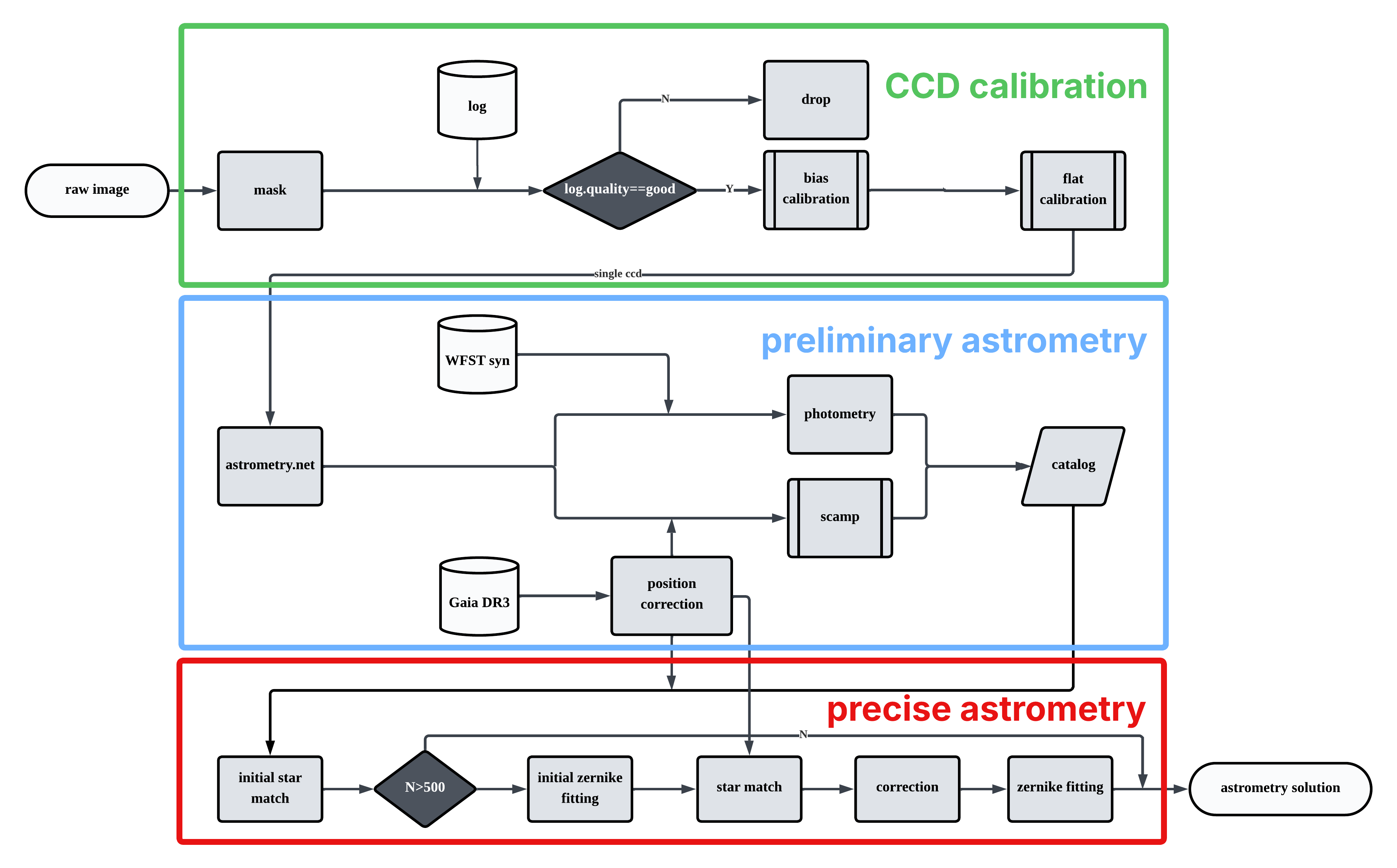}
     \caption{Framework of global GD modeling. The green box represents the CCD calibration module, which is used to remove instrumental signals from the images. The blue box denotes the initial astrometric calibration module for extracting source catalogs from the images and provides the CD matrix together with CRVAL and CRPIX values for GD modeling. The red box indicates the precise astrometric calibration module, which outputs a combined source catalog for all CCDs after applying the GD model and corrections.}
     \label{fig:flow}
\end{figure*}

\subsection{Global GD model construction}
\label{sec:wcs}

Before constructing GD model, we introduce 24 free parameters to characterize the relative positions of the mosaic CCDs. As shown in \figref{fov}, the focal-plane coordinate system is anchored at the central CCD. For a given $CCD_i$, two translation parameters $(\delta x_i, \delta y_i)$ and one rotation parameter $\alpha_i$ are defined to describe its displacement and orientation relative to the central CCD. The stellar position $(X_{ij}, Y_{ij})$ on the focal plane is expressed as

\begin{equation}
    \left ( \begin{matrix}
        X_{ij} \\
        Y_{ij} \\
        \end{matrix} \right ) = 
    \left ( \begin{matrix}
        \Cos \alpha_i & \Sin \alpha_i \\
        -\Sin \alpha_i & \Cos \alpha_i \\
        \end{matrix} \right ) \, 
    \left ( \begin{matrix}
        x_{ij}+\delta x_i \\
        y_{ij}+\delta y_i \\
        \end{matrix} \right ).
\end{equation}
Next, $n$th-order Zernike polynomials are applied to $(X_{ij}, Y_{ij})$ to characterize the distortions in the $X$ and $Y$ directions, resulting in the transformed coordinates
\begin{equation}
\label{eqn:transform}
    \begin{aligned}
        X_{ij}' & = X_{ij} + \Delta X = X_{ij} + \sum_{k=0}^{n} a_{X, k} Z_k (X_{ij}, Y_{ij}), \\
        Y_{ij}' & = Y_{ij} + \Delta Y = Y_{ij} + \sum_{k=0}^{n} a_{Y, k} Z_k (X_{ij}, Y_{ij}).
    \end{aligned}
\end{equation}

For the reference catalog, we adopt the CD matrix, reference coordinates (CRVAL), and reference pixel (CRPIX) recorded in the header of the FITS file of an exposure to transform the \Gaia reference positions of the matched stars into focal-plane coordinates $(X_G, Y_G)$. The $2n+24$ model parameters are then determined by minimizing the residuals between these transformed coordinates and the corresponding observed positions using a least-squares approach.

\figref{flow} presents a flowchart summarizing the overall framework. In the \code{precise astrometry} module, the star-matching and Zernike-fitting steps are iterated once to recover stars that are discarded in the initial matching (based on the astrometry solution from a single CCD). 

\begin{figure*}[ht]
     \center
     \includegraphics[width=\textwidth]{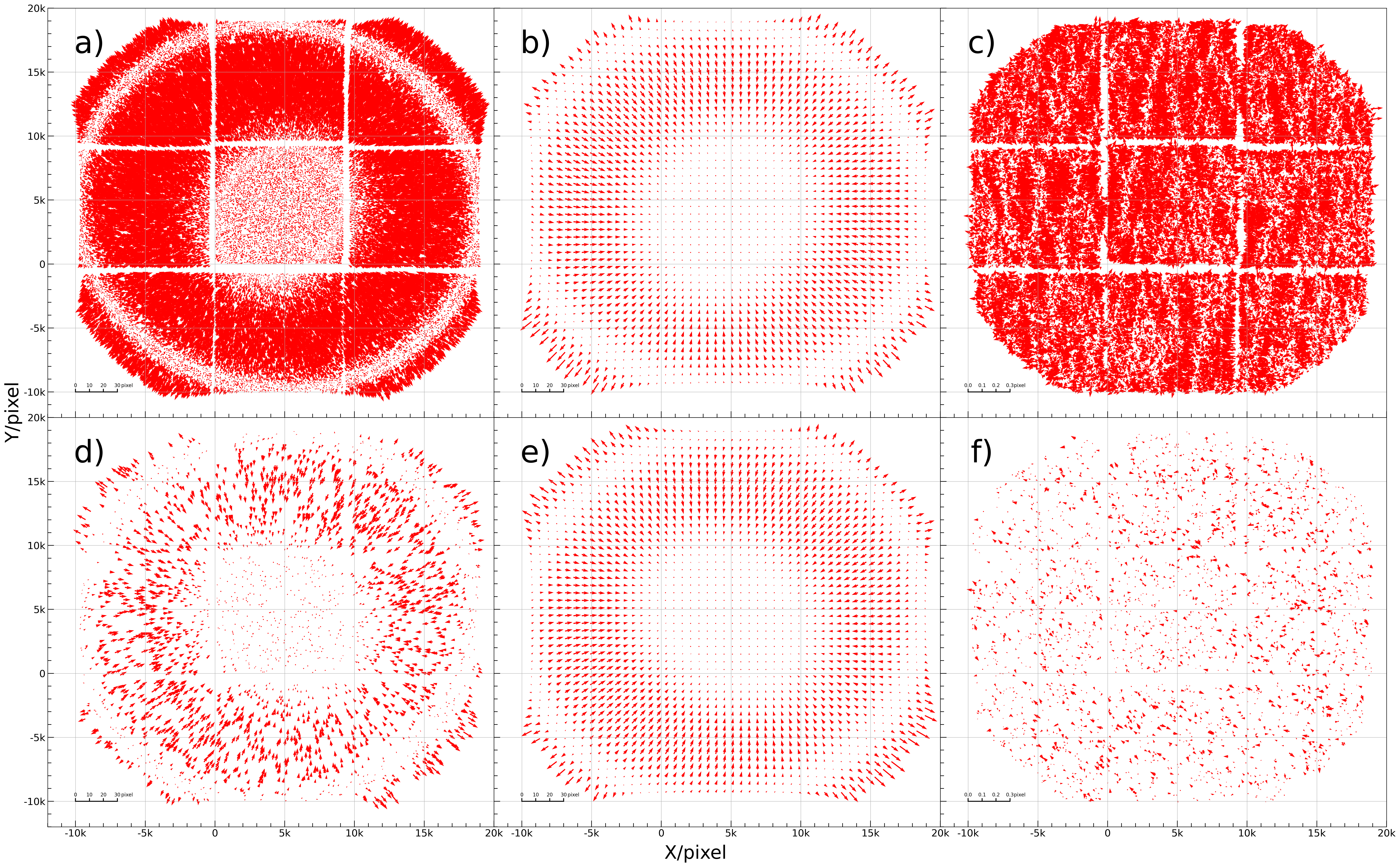}
     \caption{Left to right shows the position offsets of the matched stars, the model global GDs with Zernike polynomicals, and the residuals between the observed offsets and the model GD, respectively. The upper panels correspond to a dense field, while the lower panels correspond to a sparse field. Arrow lengths are scaled according to the reference scale bar shown in each panel.}
     \label{fig:distortion}
\end{figure*}

\begin{figure*}[ht]
     \center
     \includegraphics[width=0.9\textwidth]{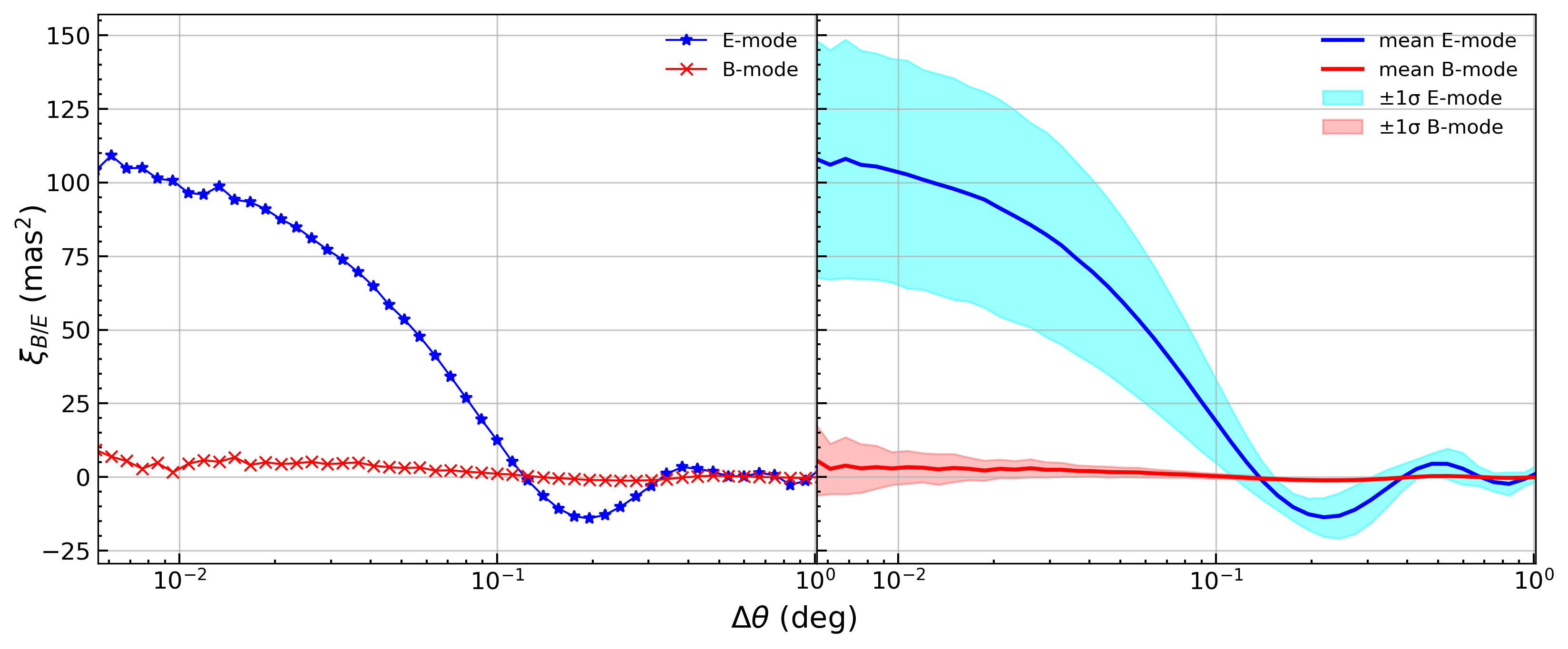}
     \caption{Typical E- and B-mode correlation functions. The left panel shows the correlation functions for the exposure corresponding to panel~(c) of \figref{distortion}, while the right panel shows the correlation functions averaged over all exposures in the 20250218 dataset. The solid curves denote the mean values, and the shaded regions indicate the $1\sigma$ scatter among the exposures.}
     \label{fig:be_mode}
\end{figure*}

\begin{figure*}[ht]
     \center
     \includegraphics[width=\textwidth]{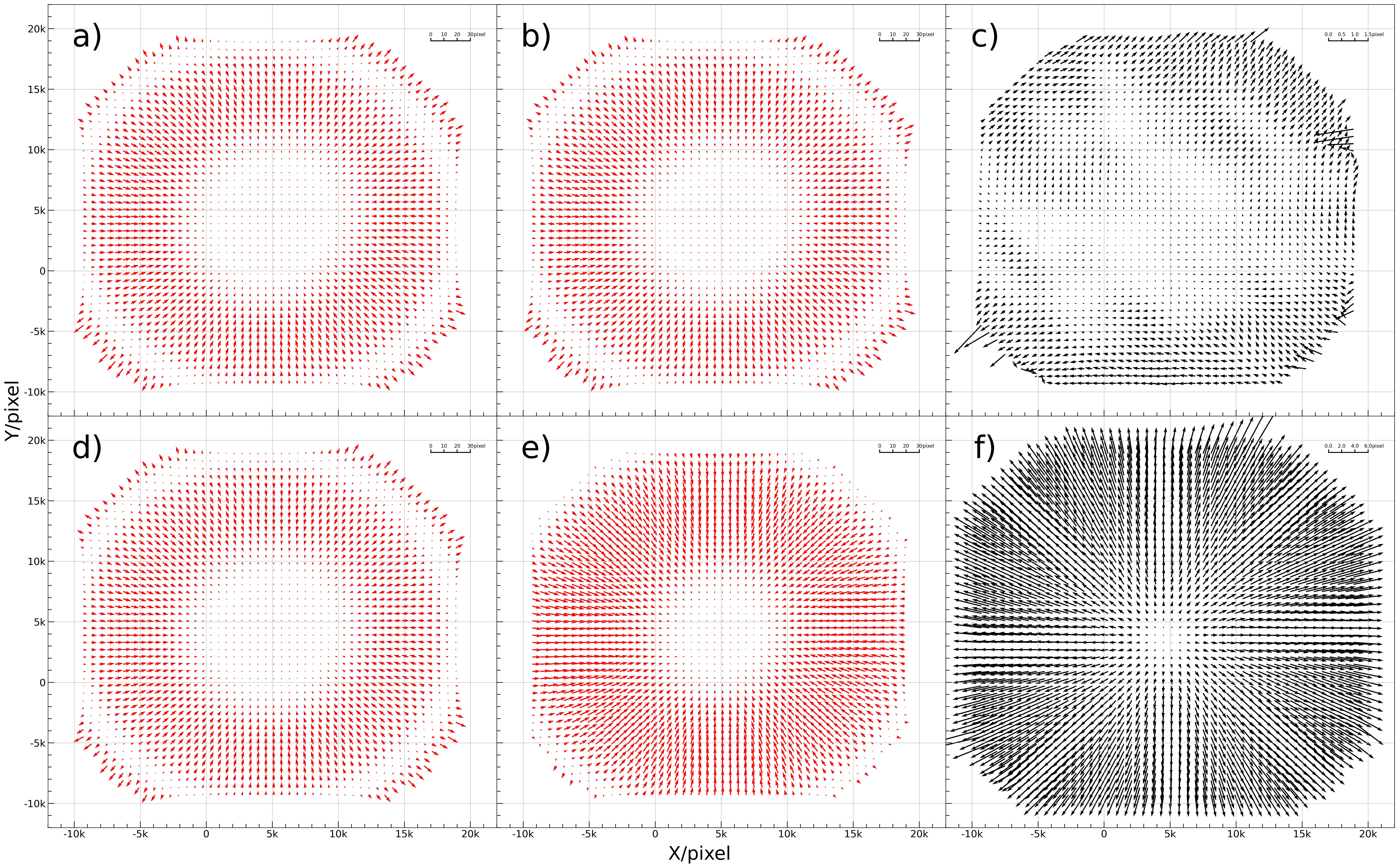}
     \caption{Global GD models and their difference maps for two pairs of consecutive exposures in the 2016HO3 field. Panels~(a) and (b) show similar GD models in one pair of consecutive exposures, with panel~(c) presenting their difference. Panels~(d) and (e) display significantly different GD models in another pair of exposures, with panel~(f) showing their difference.}
     \label{fig:stab}
\end{figure*}

\begin{figure}[h]
     \center
     \includegraphics[width=\columnwidth]{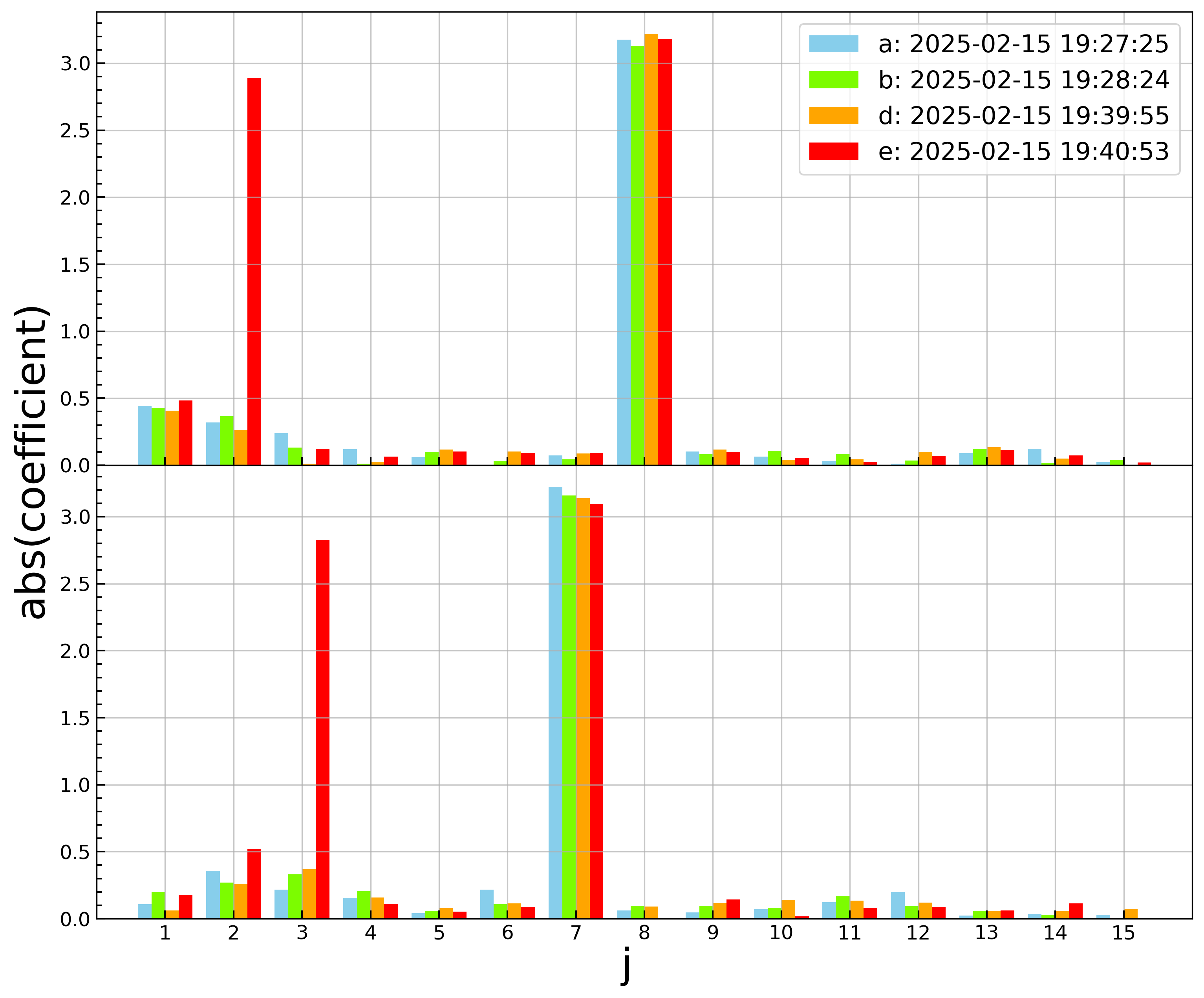}
     \caption{Absolute values of the first 15 zernike coefficients for panels~(a), (b), (d) and (e) in \figref{stab}. The upper panel shows these coefficients along $X$ axis, while lower panel shows those along the $Y$ axis.}
     \label{fig:power}
\end{figure}

\begin{figure*}[t]
     \center
     \includegraphics[width=\textwidth]{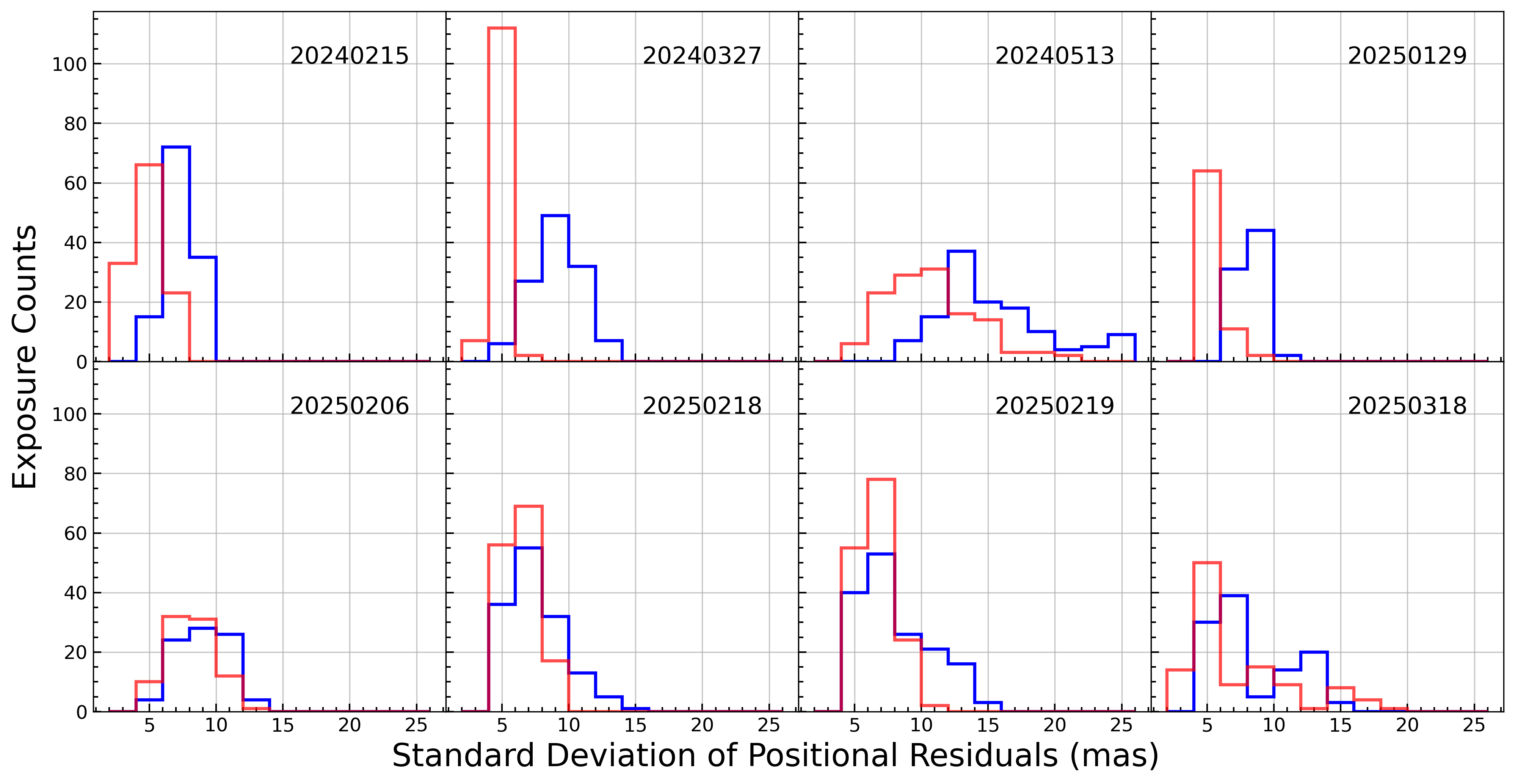}
     \caption{Histograms of the standard deviations of GD-corrected residuals for all exposures within a single night. The blue bars represent the standard deviations along the $X$ axis, while the red bars represent those along the $Y$ axis.}
     \label{fig:resid}
\end{figure*}

\subsection{Computational Performance}
\label{sec:performance}

This pipline is executed on a compute node equipped with dual AMD EPYC 7763 processors (128 cores in total) and 1\,TB of DDR4 memory, while each exposure is processed on an independent thread. Using Zernike polynomials up to order $n=100$, the \code{precise astrometry} module takes $\sim$50\,s to deliver the astrometric calibration results for a sparse field (\roughly $10^3$ matched stars) and $\sim$300\,s for a dense field (\roughly $10^4$ matched stars). With simultaneous multithreading enabled, the system provides 256 threads, allowing all exposures from a whole night to be processed within half an hour.

\section{Results and Discussion}
\label{sec:result}

\subsection{Distortion Model}
\label{sec:distortion}

In contrast to their application in active optics systems, where Zernike polynomials are employed starting from the fourth order \citep{Megias2024}, we retain the first three orders in our GD model. These low-order modes account for the zero-point offset and the first-order affine transformations, including the global rotation and plate-scale of the focal-plane coordinate system. As a result, the global translation and rotation between the CCD focal-plane coordinate system and the astrometric reference frame is determined in a self-consistent manner, rather than being measured from individual CCDs.

We first perform an initial global GD modeling of all observations using Zernike polynomials up to the 100th order. For the final global GD modeling, Zernike polynomials up to the 300th order are applied to further improve the modeling accuracy. \figref{distortion} shows the GD modeling procedure for two representative 30\,s exposures: one from a dense stellar field in the wide field survey (WFS) region (33,465 matched stars, corresponding to \roughly1.55\,arcmin$^{-2}$), and the other from a sparse stellar field (2,537 matched stars, \roughly0.12\,arcmin$^{-2}$). Panels~(b) and (e) in \figref{distortion} reveal that the GD of WFST exhibits a large-scale radial pattern: the distortion increases from the field center outward, then decreases, and rises again toward the edges, with a maximum position offset of \roughly10\,pixels. Such large-scale GD pattern is left in the optical design of the telescope. In other words, different telescopes may display different GD patterns. Nevertheless, theoretically speaking, these GD patterns can be effectively modeled using Zernike polynomials. Furthermore, as long as the matched stars are distributed across all mosaic CCDs (panels~(a) and (d) in \figref{distortion}), this method can still deliver a robust global GD model, even when the matched stellar density within the FoV is as low as $\sim0.1$\,arcmin$^{-2}$.

Panels~(c) and (f) in \figref{distortion} present the positional residuals of the two stellar fields after correcting for the GD, with standard deviations of 5-8 mas along a single axis. Panel~(c), in particular, reveals a pronounced spatial pattern in the residuals. A comparison of consecutive exposures further reveals that this pattern evolves on short timescales. Similar spatial residual patterns have also been reported for DES and Hyper Suprime-Cam (HSC) data \citep{Fortino2021,Leget2021}, where they are attributed to atmospheric turbulence. Following the formalism described in the appendix of \citet{Bernstein2017}, we compute the E- and B-mode correlation functions for this exposure, as shown in the left panel of \figref{be_mode}. The E-mode correlation funtion has an amplitude of $\sim 100\mas^2$, while the B-mode is consistent with zero. This indicates that the residual field is dominated by a curl-free component, further suggesting an atmospheric origin for the observed pattern. The E- and B-mode correlation functions computed from all exposures in the 20250218 dataset (right panel of \figref{be_mode}) show a similarly strong E-mode and negligible B-mode. This suggests that, after the GD correction, the remaining astrometric residuals are strongly affected by atmospheric turbulence. Such atmospheric distortions can be modeled using two-dimensional Gaussian processes to improve astrometric accuracy, which is beyond the scope of this work.

\subsection{Distortion Stability}
\label{sec:stab}

In the night of February 15, 2024, the 2016HO3 field was observed with 59 consecutive 30\,s exposures, spanning nearly one hour with airmass of 1.11$-$1.24. Using our method, we derive the global GD models for all 59 exposures. 

\figref{stab} presents two representative examples drawn from pairs of consecutive exposures. The first example is shown in panel~(a) and (b) of \figref{stab}. These two exposures display GD patterns that are highly similar to each other, and such pattern is the most commonly observed in WFST imaging data. Their difference map (panel~(c)) reveals residuals at the level of $\sim$0.1 pixel with clear planar structure. 

By contrast, panels~(d) and (e) in \figref{stab} display a markedly different GD patterns. And the corresponding difference map (panel~(f)) shows residuals at the level of $\sim$1 pixel—an order of magnitude larger than those in panel~(c) of \figref{stab}. Unlike the irregular residual structures seen in panel~(c), the residuals in this example form an asterism-like pattern. To investigate the physical origin of this variation, the absolute values of the first 15 Zernike coefficients derived from the global GD models of the four exposures are shown in \figref{power}. In the fourth exposure, the second coefficient along the $X$ axis and the third coefficient along the $Y$ axis are markedly enhanced, indicating that tilt aberrations dominate the change in GD observed in panel~(f) of \figref{stab}. Such aberrations typically arises from tilts of optical elements. Given the short timescale of the variation and the $\sim$2.5 increase in coefficient amplitude, adjustments of the ADC are suspected as a possible cause.

Apart from this case, the global GDs in all four exposures are dominated by coma. This aberration mainly arises from the optical design of WFST, which incorporates an $F/2$ hyperbolic concave primary mirror with high-order aspheric terms, and is therefore intrinsic to all exposures.

Taken together, these two examples demonstrate that WFST GD can vary significantly even between adjacent exposures of the same field. Consequently, independent global GD modeling is required for each individual exposure.

\begin{table*}
     \setlength{\tabcolsep}{10pt}
     \begin{center}
     \footnotesize
     \caption{Relative positions of nine CCDs in the focal plane of WFST }
     \label{tab:position}
          \begin{tabular}{crrrcccccc}
          \hline
          CCD & $\overline{\delta x}$ & $\overline{\delta y}$ & $\overline{\alpha}$ & $\sigma (\delta x)$ & $\sigma (\delta y)$ & $\sigma (\alpha)$ & $\overline{\sigma_{\mathrm{model}}(\delta x)}$$^{a}$ & $\overline{\sigma_{\mathrm{model}}(\delta y)}$ & $\overline{\sigma_\mathrm{model}(\alpha)}$ \\
             & pixel & pixel & deg & pixel & pixel & $10^{-4}$\,deg & pixel & pixel & $10^{-6}$\,deg \\
          \hline
          0 & $-$9695.00 & 9447.52 & $-$0.1262 & 0.033 & 0.042 & 2.6 & 0.010 & 0.018 & 1.6 \\
          1 & 28.32 & 18943.28 & $-$90.0463 & 0.040 & 0.024 & 2.4 & 0.016 & 0.010 & 1.5 \\
          2 & 9701.97 & 9409.03 & $-$0.3208 & 0.037 & 0.022 & 2.7 & 0.010 & 0.012 & 1.8 \\
          3 & $-$9687.91 & 8960.22 & $-$90.1606 & 0.036 & 0.045 & 2.6 & 0.010 & 0.016 & 1.5 \\
          4 & 0 & 0 & 0 & --- & --- & --- & --- & --- & --- \\
          5 & 9710.62 & 8916.28 & $-$89.9867 & 0.037 & 0.014 & 2.5 & 0.010 & 0.009 & 1.5 \\
          6 & $-$9446.08 & $-$9970.71 & $-$0.1572 & 0.056 & 0.050 & 2.9 & 0.019 & 0.019 & 1.8 \\
          7 & $-$28.45 & $-$10001.73 & $-$0.0592 & 0.047 & 0.026 & 2.5 & 0.017 & 0.011 & 1.5 \\
          8 & 9374.72 & $-$10026.31 & $-$0.0941 & 0.060 & 0.027 & 3.0 & 0.017 & 0.012 & 1.6 \\
          \hline
          \end{tabular}
     \end{center}
     \begin{flushleft}
          \footnotesize \textbf{Note.} $^{a}$ Mean of the uncertainties of fitting parameters from the GD modeling.
     \end{flushleft}
     \end{table*}

\subsection{Residuals}
\label{sec:residuals}

Based on 945 $r$-band exposures, we analyze the distribution of standard deviations of astrometry residuals, as shown in \figref{resid}. With the exception of the 20240513 dataset, the median standard deviation for each dataset is below 10\mas, indicating that our method suppresses GD-induced position errors to $<0.05$\,pixel. For dataset from 20240513, the residuals after GD modeling are significantly larger, with the median standard deviations of 14.27\mas along the $X$ axis and 10.21\mas along the $Y$ axis, even though the initial stellar position offsets and the distortion model for each exposure show no apparent anomalies. Residual maps from that night further reveal clear spatial structures, with some exposures even showing vortex-like patterns. Considering the relatively long timescale of these effects, we argue that atmospheric variations within a single night (e.g., strong turbulence in the upper atmosphere) are the most likely cause.

In addition, except for the datasets from 20250206, 20250218, and 20250219, all other datasets show systematic offsets in the median standard deviations along the two orthogonal coordinate axes. Although WFST is equipped with an ADC system, incomplete correction results in residual chromatic dispersion effect, which gives rise to these systematic offsets. Further discussion is provided in \secref{chromatic}.

\begin{figure*}[t]
     \center
     \includegraphics[width=0.95\textwidth]{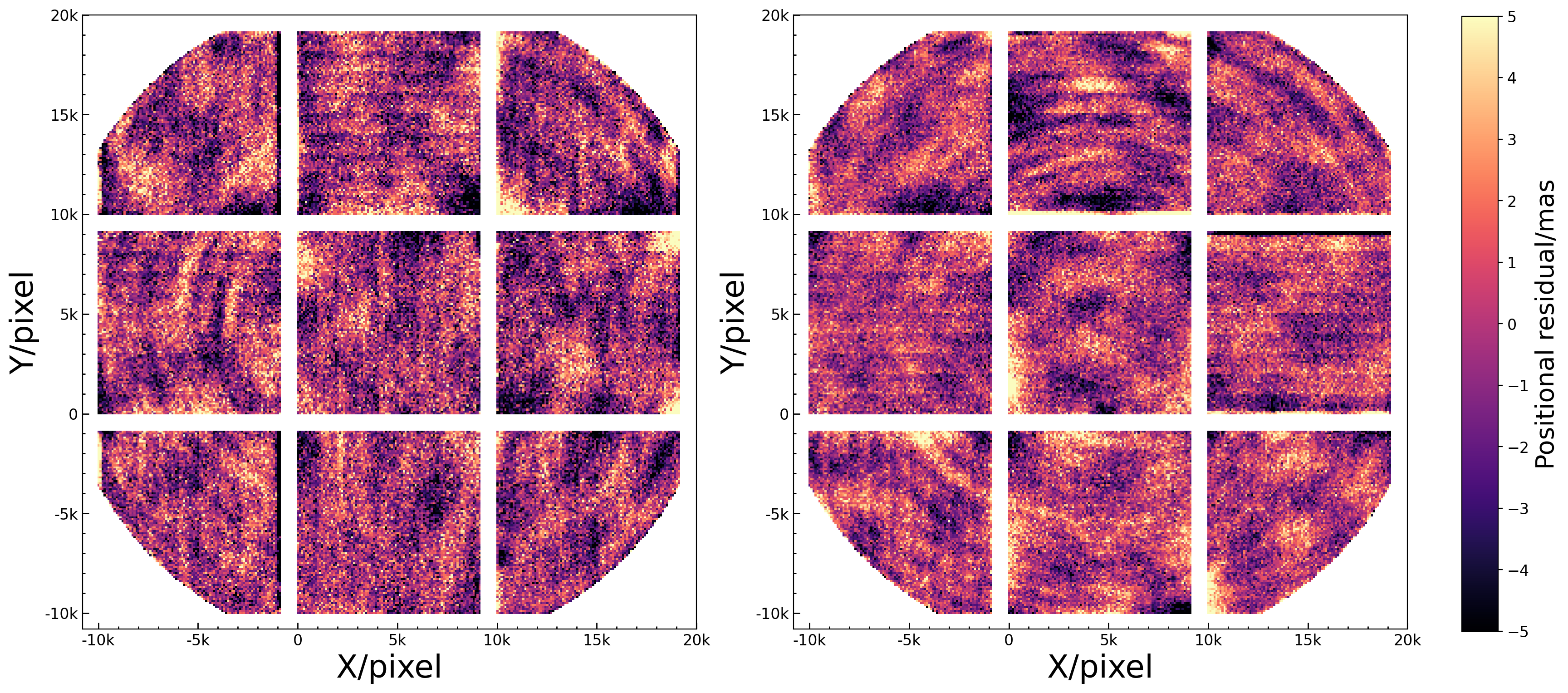}
     \caption{Stacked maps of positional residuals for each CCD after GD modeling. The left panel shows the mean residual in the X-direction, while the right panel shows the mean residual in the Y-direction. Each pixel represents the mean positional residual averaged within a $100 \times 100$ pixel bin in the local CCD coordinate system.}
     \label{fig:sensor}
\end{figure*}

\figref{sensor} shows the stacked maps of positional residuals for each CCD after GD modeling. Persistent residual patterns appear near the edges and corners of the CCDs, arising from detector-induced systematic astrometric uncertainties. Because these features are discontinuous across the focal plane, they cannot be effectively captured by Zernike polynomials. By contrast, comparison with panel~(c) of \figref{distortion} shows that atmospheric effects remain the dominant source of astrometric errors in single-exposure calibration.

\begin{figure}[h]
     \center
     \includegraphics[width=\columnwidth]{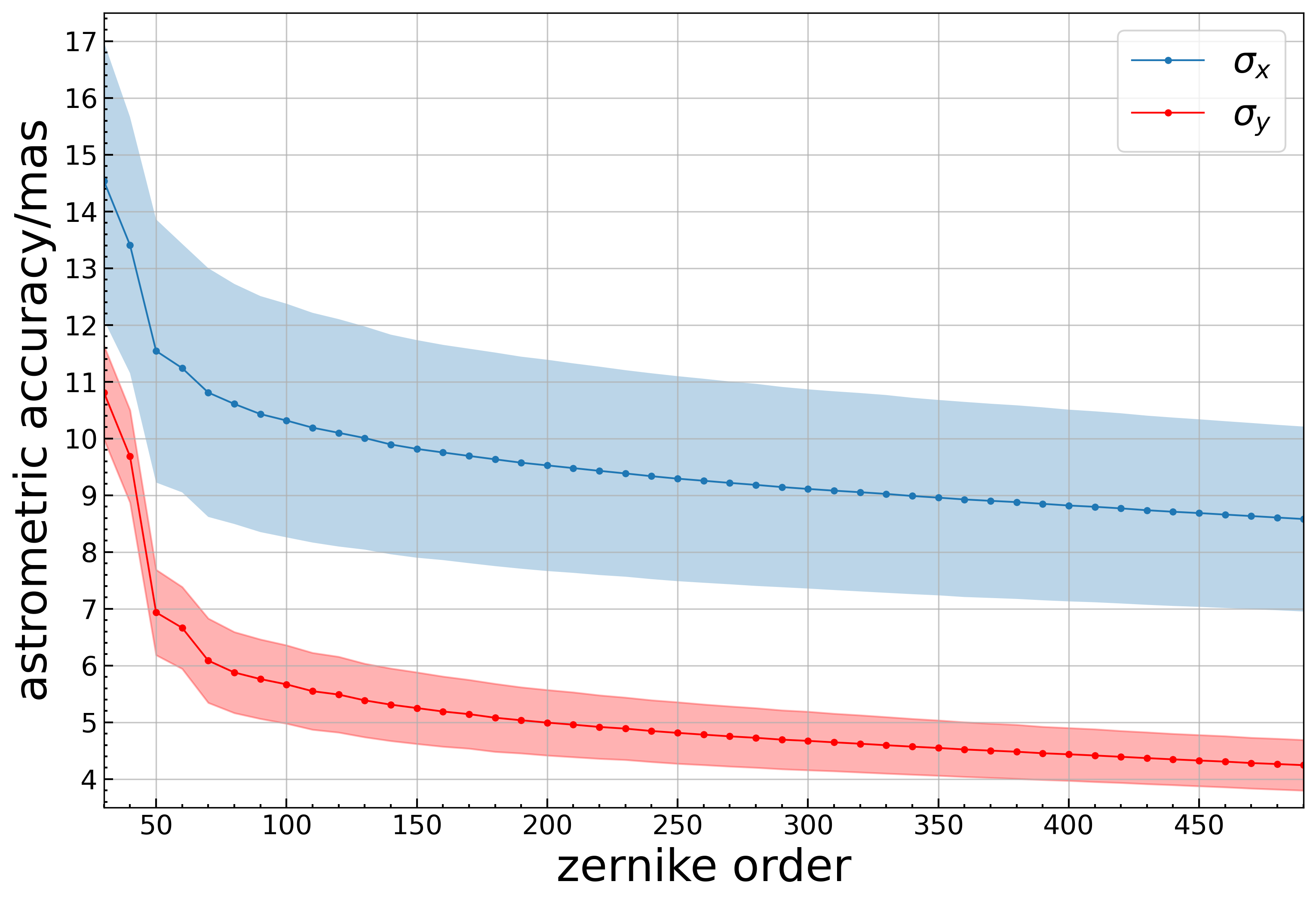}
     \caption{Variation of astrometric accuracy for all exposures in the 20240327 dataset after correction with Zernike polynomials of different orders. The solid lines indicate the mean astrometric accuracy across all exposures for each order, while the shaded regions represent the 1$\sigma$ scatter among these exposures.}
     \label{fig:order}
\end{figure}

\subsection{Relative Positions of CCDs}
\label{sec:position}

Another important capability of this method is the determination of the relative positions of the CCDs on the two-dimensional focal plane for each observation. From the 20250218 dataset, 82 exposures with more than 10,000 matched stars each are selected, and the relative position parameters ($\delta x_{i}$, $\delta y_{i}$, $\alpha_{i}$) for each CCD are statistically analyzed, as summarized in \tabref{position}. Comparison of the average relative CCD positions with our engineering design specifications reveals pixel-level discrepancies, demonstrating that digital stitching based on the design drawings is inadequate for milliarcsecond-level astrometric calibration.

Furthermore, for individual exposures, the uncertainties in determining the relative position parameters are below 0.02\,pixel for translations and below 7.2\mas for rotations. And notably, across multiple exposures within a single observing night, the standard deviation of the relative position parameters is larger. This variation may arise from changes in telescope attitude, camera rotation angle, or thermal expansion effects associated with temperature fluctuations during telescope movement. Nevertheless, our results demonstrate that modeling GD with Zernike Polynomials provides a robust solution for determining the relative positions of CCDs, achieving a precision of 0.2\unit{\mu m}.

\subsection{Chromatic Dispersion Effects}
\label{sec:chromatic}

The correction processes in \figref{flow} stand for some chromatic dispersion effects correction, primarily DCR and lateral color. Both effects produce systematic position offsets that depend on the spectral energy distribution of source: DCR effects arises from the wavelength-dependent refractive index of the atmosphere \citep{Lin2020}, whereas lateral color effect results from the wavelength-dependent refractive indices of the optical elements. Correction for these two effects is essential for obtaining high-accuracy position of a star \citep{Anderson2006}. However, in this paper we neglect them for two reasons. First, extremely red or blue stars constitute only a small fraction of the matched stars, and therefore have a negligible impact on the global GD modeling. Second, chromatic dispersion effects are more significant in shorter-wavelength bands under given atmospheric temperature and pressure conditions \citep{Stone2002, Sullivan2015}. In the $r$ band, position offsets ratio caused by chromatic dispersion effects are expected to be below 20\masmag \citep{Bernstein2017}. Moreover, among the achieved observations of different bands, the $r$-band exposures provide images containing more than 500 unsaturated stars in a single exposure for reliable global GD model construction.

At an observing site with fixed atmospheric pressure and temperature, the DCR effect theoretically depends only on the zenith distance (\ZD; \citealt{White2022}), whereas the lateral color effect is determined by the optical design of a telescope and usually fixed for observations by the same telescope  \citep{Bernstein2017}. Because WFST is equipped with an ADC system, classical chromatic dispersion is partially mitigated, while the remaining residuals may vary with the operating state of the ADC. Furthermore, assessing the impact of these chromatic dispersion effects on astrometric calibration requires a sufficiently large sample of stars within the FoV to ensure adequate color coverage. For our regular WFST observations with a fixed exposure time (mostly 60\,s), the number of detected stars in a single exposure meets the requirement with $g, r$ and $W$, but declines significantly with $u$, $i$ and $z$. A detailed correction of chromatic dispersion effects for astrometric calibration in WFST bands will be addressed in a future work.

\subsection{How Many Zernikes are Necessary?}
\label{sec:orders}
Correction with Zernike polynomials of different orders is also tested on the 20240327 dataset. As shown in \figref{order}, the astrometric accuracy improves rapidly as more free parameters are included when the order is below $\sim$70. Beyond this threshold, further increases in the Zernike order result in only marginal improvements. This behavior motivates a closer examination of what kind of effects are progressively captured as higher-order Zernike terms are included. To investigate this in a controlled manner, we turn to simulated data, for which the underlying distortion components are precisely known.

\begin{figure*}[t]
     \center
     \includegraphics[width=\textwidth]{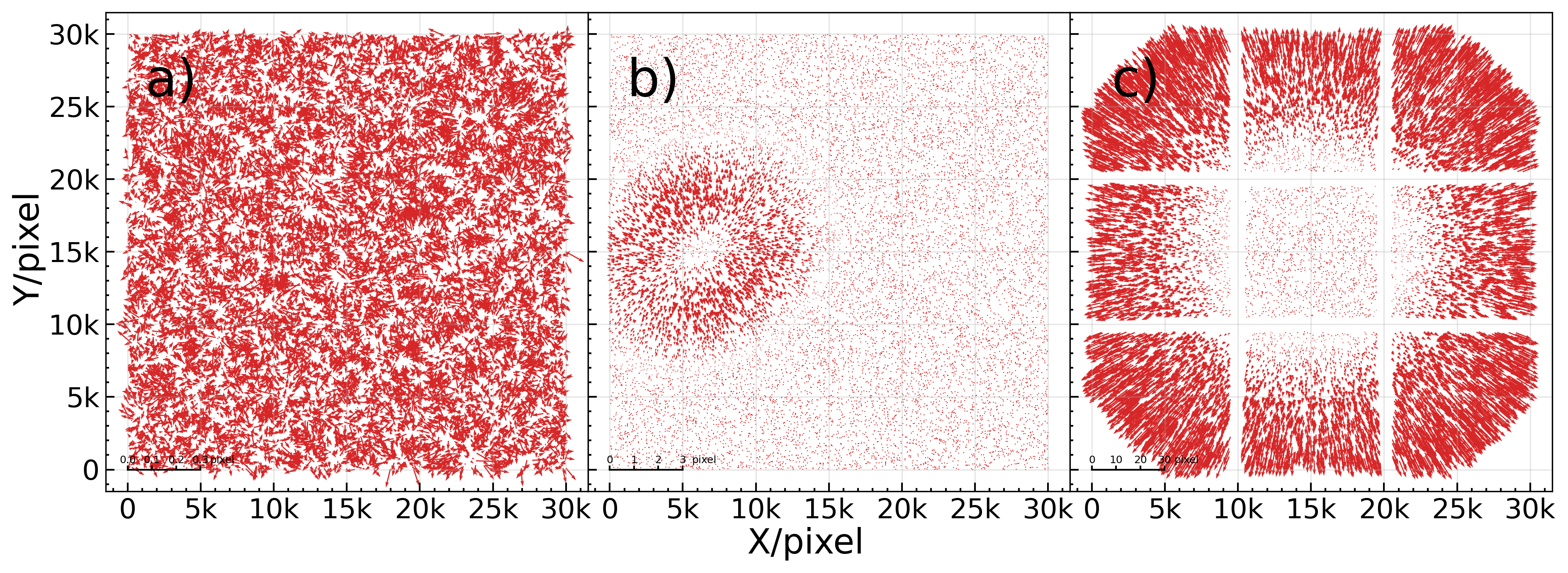}
     \caption{Simulated stellar positional offset maps. Panel~(a) illustrates the positional offsets caused by random noise and atmospheric turbulence. Panel~(b) shows the small-scale optical distortion component introduced in the simulation. Panel~(c) presents the total positional offset map resulting from all simulated effects.}
     \label{fig:mock}
\end{figure*}

In the simulations, we focus on two major contributors to astrometric offsets: optical distortion and atmospheric turbulence. We begin by sampling 15,000 points within a $30{,}000\times30{,}000$ pixel rectangular region to represent the positions of simulated stars. The initial positional offsets of these stars are drawn from a Gaussian distribution with a standard deviation of 3\mas. To model the effects of atmospheric turbulence, we generate a 2D vector field based on the von Kármán model (see Appendix~A of \citealt{Heymans2012}).  Briefly, the field is constructed from an isotropic two-dimensional power spectrum of the form
\begin{equation}
\label{eqn:von}
\footnotesize
     P(\ell) \propto \left( \ell^{2} + \frac{1}{\theta_{0}^{2}} \right)^{-11/6} ,\\[6pt]
\end{equation}
where $\theta_{0}$ represents the outer scale of the turbulence. The resulting stellar positional offset field, including both random noise $\boldsymbol{\sigma}(x_i, y_i)$ and atmospheric turbulence component $\boldsymbol{A}(x_i, y_i)$, is shown in panel~(a) of \figref{mock}. Optical distortion is simulated as the sum of two components. The large-scale distortion, denoted as $\boldsymbol{O_{1}}(x_i, y_i)$, is described by a radially symmetric function given by
\begin{equation}
\label{eqn:loptic}
\footnotesize
     \begin{cases}
         r = \sqrt[2]{{\frac{(x_i-15000)}{22000}}^2+{\frac{(y_i-15000)}{22000}}^2}, \\[6pt]
         \boldsymbol{O_{1}}(x_i, y_i) = 25 r^3 - 8 r^5, \\[6pt]
     \end{cases}
\end{equation}
where $r$ denotes the normalized distance from the field center. The small-scale optical distortion is represented by the gradient field of an anisotropic Gaussian potential centered at $(x_c, y_c)$, with principal-axis scales $(\sigma_a, \sigma_b)$ and orientation $\phi$. This component is expressed as
\begin{equation}
\label{eqn:soptic}
\footnotesize
     \begin{cases}
          \begin{pmatrix}
               x'_{i} \\ y'_{i}
          \end{pmatrix}
               =
          \begin{pmatrix}
               \cos\phi & \sin\phi \\
               -\sin\phi & \cos\phi
          \end{pmatrix}
          \begin{pmatrix}
               x_i-x_c \\ y_i-y_c
          \end{pmatrix}, \\[20pt]
          d^2 = \left(\frac{x'_i}{\sigma_a}\right)^2
          + \left(\frac{y'_i}{\sigma_b}\right)^2, \\[6pt]
          \boldsymbol{O_2}(x_i, y_i)  = A\,e^{-d^2/2}
          \begin{pmatrix}
               \cos\phi & -\sin\phi \\
               \sin\phi & \cos\phi
          \end{pmatrix}
          \begin{pmatrix}
               -\dfrac{x'_i}{\sigma_a^2} \\[6pt]
               -\dfrac{y'_i}{\sigma_b^2}
          \end{pmatrix}, \\
     \end{cases}
\end{equation}
where the amplitude $A$ is approximately an order of magnitude smaller than that of the large-scale component. The resulting positional offsets associated with this small-scale optical distortion are shown in panel~(b) of \figref{mock}.

\begin{figure*}[t]
     \center
     \includegraphics[width=0.85\textwidth]{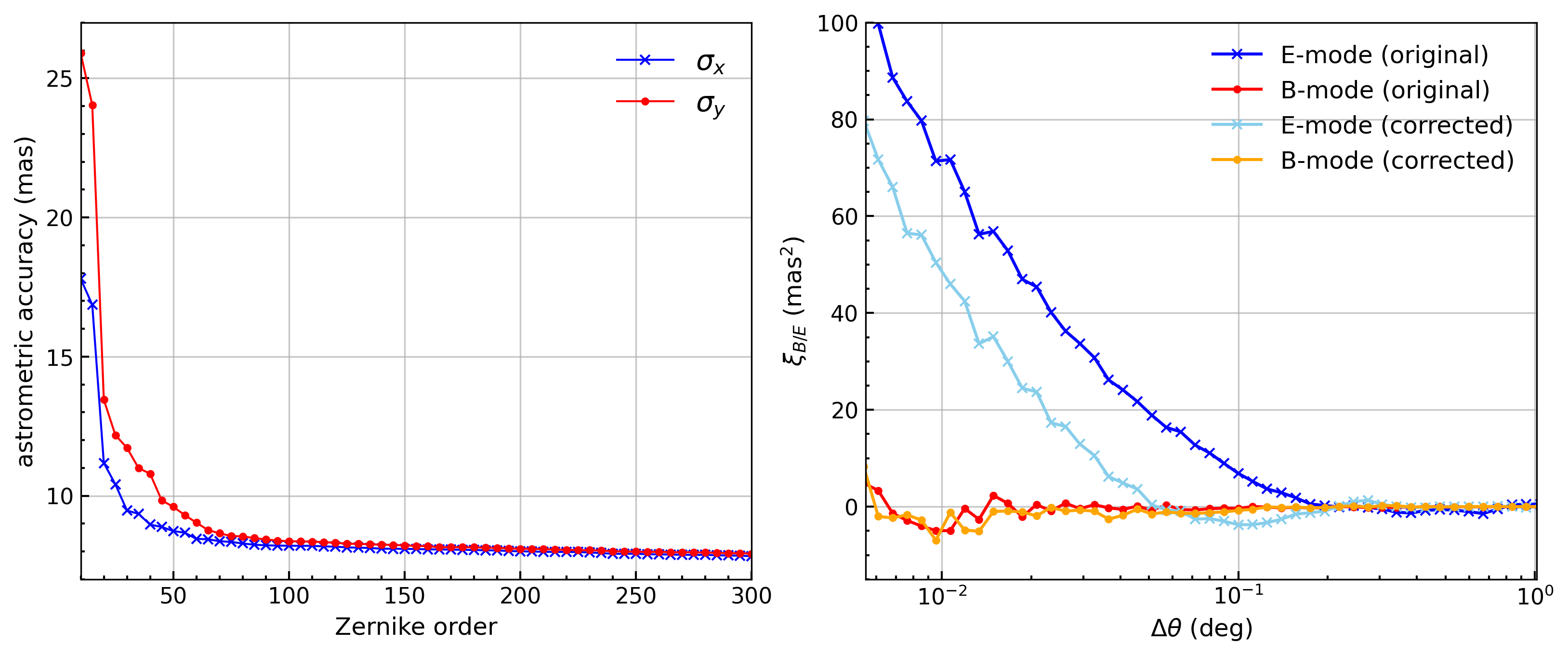}
     \caption{The left panel shows the astrometric accuracy of the simulated data after modeling with Zernike polynomials of different orders. The right panel presents the E- and B-mode correlation functions of the stellar positional offsets induced by atmospheric turbulence in the simulation, together with the E- and B-mode correlation functions of the residual stellar positional offsets after GD correction using 300th-order Zernike polynomials.}
     \label{fig:mock_order}
\end{figure*}

\begin{figure*}[t]
     \center
     \includegraphics[width=\textwidth]{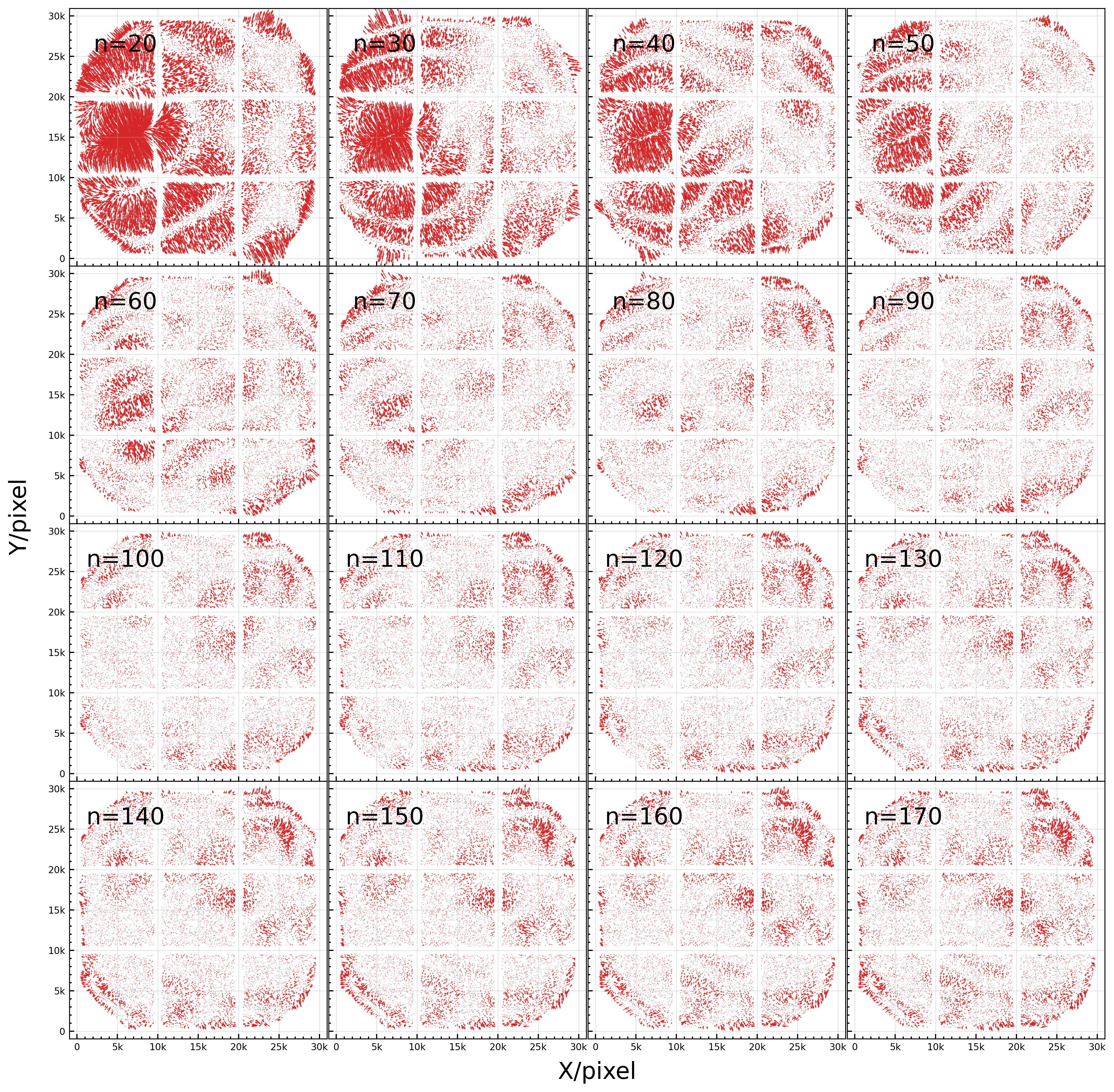}
     \caption{Residual maps between the GD models derived using Zernike polynomials of orders from 20 to 170 and the input optical distortion model in the simulation. The vector scale is identical to that used in panel~(a) of \figref{mock}.}
     \label{fig:mock_resid}
\end{figure*}

By combining these effects and selecting stars that would be detectable by nine square CCD within the rectangular field (panel~(c) of \figref{mock}), we obtain the final simulated stellar positional offset field $\boldsymbol{u}(x_i, y_i)$, which can be written as
\begin{equation}
\label{eqn:mdistortion}
\footnotesize
     \boldsymbol{u}(x_i, y_i) = \boldsymbol{O_1}+\boldsymbol{O_2}+\boldsymbol{A}+\boldsymbol{\sigma}. \\ [6pt]
\end{equation}
This field is then modeled using Zernike polynomials ranging from 20 to 300. The standard deviation of the residuals as a function of Zernike order is shown in the left panel of \figref{mock_order}. The residuals between the fitted GD models and the known optical distortion field are presented in \figref{mock_resid}. These results demonstrate that small-scale optical distortions are successfully captured once the Zernike order reaches $\sim$100. At the same time, when using the Zernike polynomials exceeds $\sim$50th order, the model begins to absorb relative large-scale components from the atmospheric turbulence field. This behavior is also further illustrated in the right panel of \figref{mock_order}: compared to the original atmospheric E-mode signal, the residual obtained after fitting with 300th-order Zernike polynomials exhibits both a reduced amplitude and a shorter correlation length.

The simulation results indicate that higher-order Zernike polynomials tend to capture GDs on smaller spatial scales. These distortions may originate from a combination of optical aberrations, atmospheric effects, and CCD-scale variations induced by temperature fluctuations. Consequently, within our astrometric calibration framework, the choice of Zernike order is governed by both the spatial scales of GDs and the required level of astrometric accuracy. In addition, given that this method will be applied to the full survey data in the future, adopting Zernike polynomials of 100th order for the initial fitting and 300th order for the final fitting represents a reasonable compromise between astrometric accuracy and computational cost. \\[8pt]

\subsection{Zernike Polynomials versus Bivariate Polynomials}
\label{sec:versus}

Bivariate polynomials in $x$ and $y$ are widely used in the construction of astrometry solutions, with both the PV and SIP conventions employing such expansions to describe nonlinear distortion terms \citep{Calabretta2002, Shupe2005}. For our GD modeling, we also test replacing \eqnref{transform} with 

\begin{equation}
\label{eqn:ntransform}
\footnotesize
\begin{cases}
                  X_{ij}' & = X_{ij} + \Delta X = X_{ij} + \sum_{p=0}^{n} \sum_{q=0}^{n-p} b_{X, pq} X_{ij}^{p} Y_{ij}^{q}, \\[6pt]
                  Y_{ij}' & = Y_{ij} + \Delta Y = Y_{ij} + \sum_{p=0}^{n} \sum_{q=0}^{n-p} b_{Y, pq} X_{ij}^{p} Y_{ij}^{q}. \\[6pt]
\end{cases}
\end{equation}

\noindent The results show that, for WFST, the modeling accuracy of GD is effectively the same ($\ll 1\mas$) when using the same number of coefficients (e.g., 20th-order xy polynomials versus 231st-order Zernike polynomials). Although the accuracies are comparable, two points are worth emphasizing. First, Zernike polynomials are orthogonal, so each order represents independent information, whereas the terms of bivariate polynomials in $x$ and $y$ are not orthogonal, which can increase the instability of the fitted coefficients and lead to some problems (e.g., Runge's phenomenon, \citealt{James1987}). Second, since most FoVs of telescope are circular, different aberration modes can be independently represented by Zernike polynomials. This makes them particularly effective for diagnosing and optimizing telescope optics. In summary, although both approaches achieve similar modeling accuracy, Zernike polynomials offer a more stable and physically interpretable framework for constructing geometric distortion models.

\section{Summary}
\label{sec:sum}

We present a global GD modeling framework that incorporates CCD-level detrending, preliminary astrometry for individual exposures, cross-matching detected sources with reference stars, and a joint solution for the global GD and the relative positions of CCDs of a wide-field telescope using Zernike polynomials. Benefiting from their orthogonality and direct correspondence with optical aberration modes, Zernike polynomials not only enable robust GD modeling in fields with stellar densities as low as \roughly0.12\,arcmin$^{-2}$, but also provide a powerful tool for diagnosing and optimizing telescope optics---an advantage not achievable with traditional bivariate polynomials. In addition, the method places all CCDs into a single rigid focal-plane coordinate system. Within this unified frame, the global translation and rotation of the focal plane are naturally captured by the lowest-order Zernike terms and are determined self-consistently by the GD model.

A total of 945 $r$-band exposures from WFST, obtained over eight nights, are used to construct global GD models within this framework. We find that the median standard deviation of position offsets after correction is below 10\,mas. The global GD of WFST exhibits a characteristic radial pattern---expanding from the field center, contracting, and then expanding again. And statistical analysis of the Zernike coefficients indicates that the global GD of WFST is primarily dominated by coma, which is intrinsically linked to the optical design of telescope. Continuous observations of the 2016HO3 field further reveal that the GD is not stable: small variations at $\sim$0.1\,pixel level occur between common modes, and in some cases deviations as large as $\sim$1\,pixel are observed. Analysis of the Zernike coefficients indicates that the latter variations are mainly associated with tilt aberrations. Given the short timescales and the amplitude variations (\roughly 2.5) of coefficient, adjustments of the ADC are suspected to be a possible cause. In addition, analysis of 82 exposures from 20250218 with more than 10,000 matched stars each, demonstrates that our method achieves a precision of 0.02\,pixel (translation) and 7.2\mas (rotation) in measuring the relative CCD positions. Moreover, variations in relative CCD positions across different exposures within a single night are present, with standard deviations less than 0.1\,pixel and 1\farcs8.

Through simulation, we show that, for WFST-like telescopes, increasing the Zernike order in GD modeling progressively captures geometric distortions on smaller spatial scales. And when the Zernike order exceeds $\sim$50, the model begins to absorb the large-scale components from atmospheric turbulence in the simulated data.

This method, which can derive the global GD from single exposures, is particularly useful for images taken from wide-field telescopes equipped with mosaic CCDs (e.g., LSST and China Space Station Survey Telescope). Building on this framework, the analysis of chromatic dispersion effects for such telescopes across different bands can be naturally extended. In upcoming works of this series, we will extend this analysis to all WFST bands to futher improve astrometric calibration. Moreover, we plan to model the atmospheric turbulence in the residual maps using Gaussian process interpolation to make higher-accuracy astrometric calibration for both WFST images and catalogs.

\section*{Acknowledgments}
This work is supported by the National Key Research and Development Program of China (2023YFA1608100), the National Science Foundation of China (NSFC, Grant No. 12233005), the China Manned Space Program with grants nos. CMS-CSST-2025-A08 and CMS-CSST-2025-A20, and  Office of Science and Technology, Shanghai Municipal Government (grant Nos. 24DX1400100, ZJ2023-ZD-001).

\software{\code{astropy} \citep{Astropy2013,Astropy2018},  
          \SExtractor \citep{Bertin1996},
          \PSFEx \citep{Bertin2013}, 
          \scamp \citep{Bertin2006},
          \code{scipy} \citep{Virtanen2020},
          \code{numpy} \citep{Van2011},
          \code{astrometry.net}
          }

\bibliography{ms}{}
\bibliographystyle{aasjournal}

\end{document}